\documentclass[aps,pra,twocolumn]{revtex4}
\usepackage[dvips]{graphics}
\usepackage{graphicx}
\begin{document}
%
\title{From a single- to a double-well Penning trap}
\author{G. Ciaramicoli, I. Marzoli, and P. Tombesi}
\affiliation{Scuola di Scienze e Tecnologie,
             Universit\`a degli Studi di
             Camerino, 62032 Camerino, Italy}
\date{\today}
\begin{abstract}
The new generation of planar Penning traps promises to be
a flexible and versatile tool for quantum information studies.
Here, we propose a fully controllable and reversible way to change
the typical trapping harmonic potential into a double-well potential,
in the axial direction.
In this configuration a trapped particle
can perform coherent oscillations between the two wells.
The tunneling rate, which depends on the barrier height and width,
can be adjusted at will by varying the potential difference applied to
the trap electrodes. Most notably, tunneling rates in the range
of kHz are achievable even with a trap size of the order of
100 $\mu$m.
\end{abstract}
%
\maketitle
The challenging goal of using single electrons in Penning traps
for quantum information \cite{marzoli,ciaramicoli1,ciaramicoli6,ciaramicoli8,
pedersen,lamata}
has motivated intense research towards
a completely new generation of devices. The so-called planar Penning
traps \cite{stahl} are specifically designed to meet the demands
of quantum computation, thus
allowing for scalability as well as improved addressability of the trapped
particles.
Moreover, new pixel microstructures promise to generate complex electric
potentials, suitable both for particle transport and trapping in
racetrack and artificial crystal configurations \cite{hellwig}.
The first planar
Penning traps were operated in Mainz \cite{galve1,galve2}
and in Ulm \cite{bushev1} both at
room temperature and in a cryogenic environment. However, the elusive
goal of a single trapped electron has not yet been achieved.
The major obstacle is the
anharmonicity in the axial potential, which prevents the detection
of a single electron. To overcome this difficulty new theoretical
studies \cite{goldman} have carefully analyzed the geometry of
planar Penning traps,
with the aim of optimizing the harmonicity of the axial potential.
The results are extremely encouraging and may lead to the first
experimental demonstration of a single electron in a planar Penning
trap.

Here we demonstrate the versatility of a
planar Penning trap.
In fact, it is
able to produce a smooth variation of the trapping axial potential
from a standard harmonic well into a double well.
This result is achieved by applying suitable static voltages
to the trap electrodes. The experimenter can control
both the barrier height and width, simply by adjusting the potential
difference between the electrodes.
In particular, we have simulated the behavior of an optimized
mirror-image planar Penning trap, which consists of two identical
sets of electrodes facing each other. An electron, initially trapped
in a single-well harmonic potential, can be adiabatically placed
in a double-well potential.
Depending on the energy of the axial motion, the particle wave
function may spread over for several microns, through the barrier
between the two wells. Therefore,
the particle motional state could be prepared in a superposition of
\emph{left} and \emph{right}, with the particle being, at the
same time, in both wells.
Hence, if a single electron trapped in
a Penning trap forms a so-called \emph{geonium} atom
\cite{brown}, the electron
in the double-well potential can mimic a giant molecule
along the lines of Ref.~\cite{lesanovsky}.

Moreover, this setup
opens new perspectives for quantum information studies
with a single trapped particle.
Indeed, the position of the
electron in the double-well potential may serve as a further
qubit, encoding the $|0\rangle$ ($|1\rangle$) logical state
in the left (right) well.
Finally, the observation of a single particle, which performs
Rabi oscillations between the two wells, could find application
in interferometric schemes and precision sensing, as
proposed for trapped ions in double-well potentials
\cite{retzker}.

A typical example of double-well potential
is provided by the polynomial function
\begin{equation}\label{funct}
U(z)=a z^4-b z^2
\end{equation}
with $a$, $b > 0$.
In this case the distance between the two minima is
$L=2\sqrt{b/(2a)}$, whereas the barrier height is $E_b=b^2/(4a)$.
\begin{figure}
  \includegraphics[width=\columnwidth]{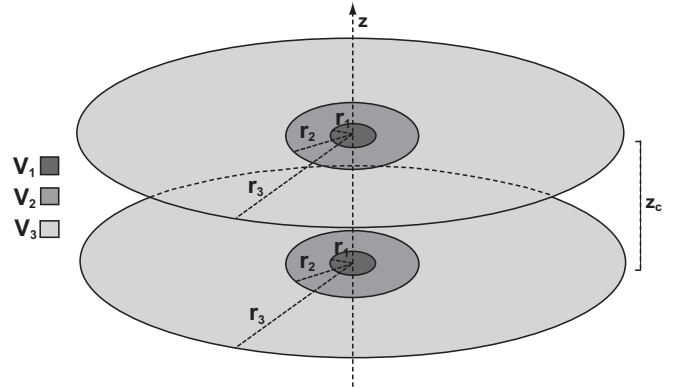}
  \caption{\label{planar} 
Schematic drawing of a mirror-image planar trap.
Each electrode plane consists of a round central electrode with radius
$r_1$ and two concentric ring electrodes with outer radii $r_2$ and $r_3$.
The distance between the electrode planes is $z_c$.}
\end{figure}
We propose to realize such a
double-well potential by means of a
mirror-image planar Penning trap \cite{goldman}.
This device consists of
a spatially uniform magnetic field, directed along
the $z$ axis (axial direction), superimposed to
an electrostatic field produced by two planar electrode structures, identically
biased and facing each other.
Each planar electrode arrangement is orthogonal to the $z$ axis and consists
of a circular electrode of radius $r_1$ surrounded by two concentric
rings with outer radii $r_2$ and $r_3$ (see Fig.~\ref{planar}).
The two planar electrode sets are separated by a distance $z_c$.
The $n$th electrode in each plane is held at the potential $V_n$.
We assume that the radius of the second ring electrode $r_3$ is much larger than
$r_1$, $r_2$ and than the distance $z_c$ between the two electrode
sets.
The trapping potential, along the $z$ axis, can be analytically calculated
in the limit $r_3 \rightarrow \infty$.
To this end, we set the origin of the $z$ axis on the
surface of the lower electrode plane and introduce the dimensionless
parameters
$\tilde{z}=z/r_1$, $\tilde{z}_c=z_c/r_1$ and $\tilde{r}_i=r_i/r_1$ with
$i=1,2$.
Hence, the electrostatic potential can be written as \cite{goldman}
\begin{eqnarray} \label{pot1}
V(\tilde{z}) &=& (V_2-V_1)[\phi_1(\tilde{z})+\phi_1(\tilde{z}_c-\tilde{z})]
                 \nonumber\\
             &+& (V_3-V_2)[\phi_2(\tilde{z})+\phi_2(\tilde{z}_c-\tilde{z})]
                 +V_3 ,
\end{eqnarray}
where
\begin{equation}\label{pot2}
\phi_i(\tilde{z})=\tilde{r}_i \int_0^\infty dk \,
       \frac{\sinh[k(\tilde{z}-\tilde{z}_c)]}{\sinh(k\tilde{z}_c)} \,
       J_1(k\tilde{r}_i),
\end{equation}
with $J_1(z)$ being the Bessel function of the first kind.

\begin{figure}
  \includegraphics[width=\columnwidth]{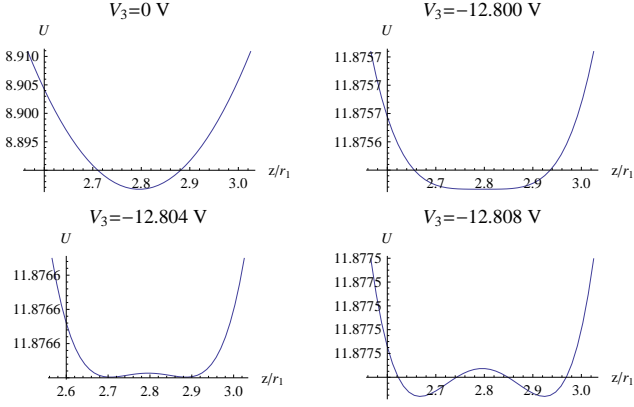} \caption{\label{trans}
Plots showing the
axial potential energy (in eV) of an electron near the center of a mirror-image
planar Penning trap.
We choose $\tilde{r}_2=4.45$, $\tilde{z}_c=5.6$, $V_1=-12.8$~V, and
$V_2=-11.4$~V
as in Ref.~\cite{goldman}.
For each plot we specify the value of the potential $V_3$ applied to the
second ring electrode.}
\end{figure}

Initially, we assume that the outermost ring electrode is grounded.
With an appropriate choice of the voltages $V_1$ and $V_2$,
the electrostatic potential takes on a parabolic shape
with the minimum
at a distance $z_c/2$ from the electrode surface
\cite{goldman}.
Now we can smoothly pass from a single harmonic trap to a double-well one,
by varying the potential $V_3$, applied to the second ring
electrode (see Fig.~\ref{trans}).
The distance between the two wells and the height of the energy barrier
are controlled simply by adjusting the potential $V_3$.

The transition from a single- to a double-well trap can be better described
by expanding the electrostatic potential, Eq.~(\ref{pot1}),
in a power series near the trap center
at $\tilde{z}=\tilde{z}_c/2$.
As a consequence, the potential energy of an electron of charge $e$
takes on
the form of Eq.~(\ref{funct}) with
\begin{eqnarray}
a&=&|e|[(V_2-V_1)a_1+(V_3-V_2)a_2], \label{a} \\
b&=&|e|[(V_2-V_1)b_1+(V_3-V_2)b_2], \label{b}
\end{eqnarray}
where
\begin{eqnarray}
a_i&=&\frac{\tilde{r}_i}{24} \int_0^\infty dk \, k^4
      \frac{J_1(k\tilde{r}_i)}{\cosh(k\tilde{z}_c/2)},\\
b_i&=&\frac{\tilde{r}_i}{2} \int_0^\infty dk \, k^2
      \frac{J_1(k\tilde{r}_i)}{\cosh(k\tilde{z}_c/2)},
\quad \mbox{with $i=1,2$}.
\end{eqnarray}
The transition from a single to a double well occurs when the
parameter $b$ changes from negative to positive.
This happens for $b=0$ when
\begin{equation} \label{V3}
 V_3= [V_2 (b_2-b_1)+V_1 b_1] / b_2.
\end{equation}
\begin{figure}
   \begin{tabular}{c}
        \includegraphics[height=3cm]{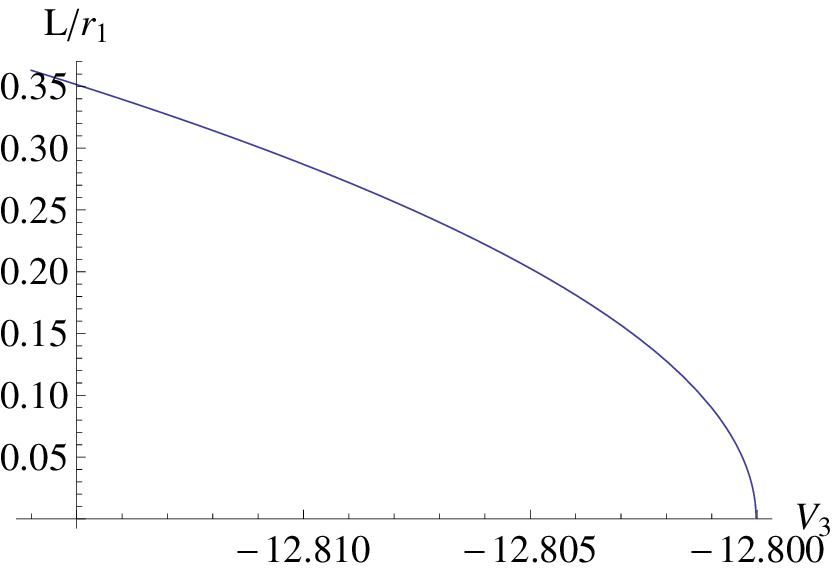} \\
        \includegraphics[height=3cm]{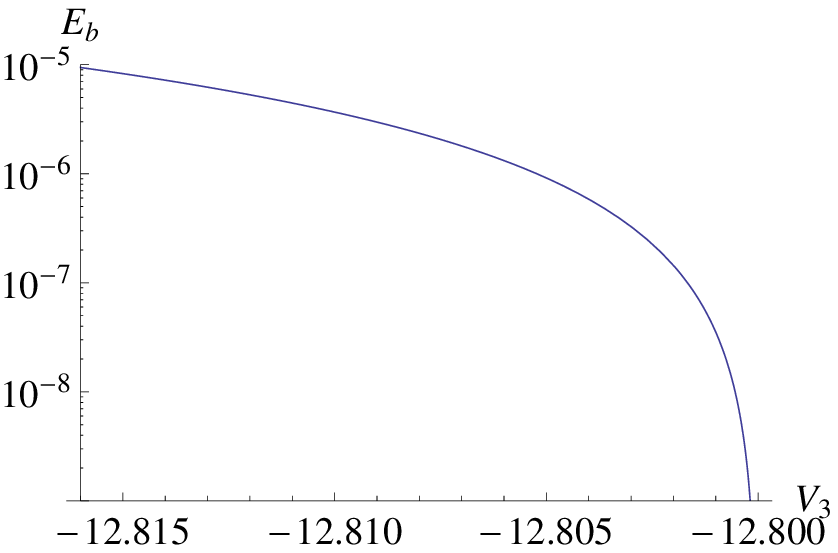}
   \end{tabular}
 \caption{\label{lv3} Top: plot of the well
distance $L$, normalized to the
the radius of the central disk electrode $r_1$,
as a function of the potential $V_3$ (in volts)
applied to the second ring electrode.
Bottom: plot of the energy barrier (in eV)
as a function of the potential $V_3$ (in volts).
The values of $\tilde{r}_2$, $\tilde{z}_c$, $V_1$ and $V_2$ are the same as in
Fig.~\ref{trans}.}
\end{figure}
Since the well distance $L$ and the energy barrier height $E_b$ are related
to the coefficients $a$ and $b$, we can calculate how these parameters
vary with the voltage $V_3$ (see Fig.~\ref{lv3}).
The distance between the two minima depends also on the radius of the
central electrode, whereas the energy barrier height does not depend on the
actual trap size.
For example, in an optimized mirror-image planar trap with $r_1=100$~$\mu$m,
$\tilde{r}_2=4.45$, $\tilde{z}_c=5.6$, $V_1=-12.8$~V, $V_2=-11.4$~V and
$V_3=-12.8013$~V
we obtain a double-well potential with minima separated by
a distance $L=10$~$\mu$m and barrier height $E_b=6 \times 10^{-8}$~eV.
In this case, the coefficients $b_1$ and $b_2$ are approximately equal and,
therefore, according to Eq.~(\ref{V3}) the transition from a single- to a
double-well trap occurs for $V_3 \simeq V_1$.

\begin{figure}
  \includegraphics[width=\columnwidth]{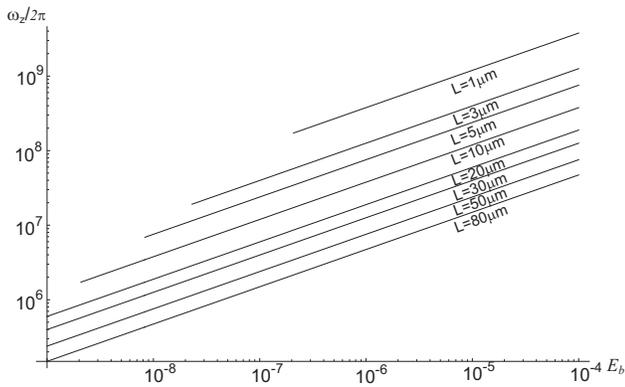}
 \caption{\label{waxial} Plot of the electron axial frequency
$\omega_z/2\pi$ (in Hz)
as a function of the barrier height (in eV) for different
choices of the well distance $L$.
For each value of $L$, there is a
minimum value of the barrier height below which
the electron motion is not confined to a single well.}
\end{figure}

The oscillation frequency of an electron, in each well, can be estimated
by expanding its potential energy, Eq. (\ref{funct}), in a power series
near the minima at $z=\pm \sqrt{b/2a}$.
Also the angular oscillation frequency depends on the width and the height
of the energy barrier
\begin{equation}
 \omega_z = \frac{4}{L} \sqrt{\frac{2 E_b}{m}} ,
\end{equation}
with $m$ being the electron mass.
The behavior of the axial frequency $\omega_z/2\pi$
is plotted in Fig.~\ref{waxial}
as a function of the
barrier height, for different values of the well distance.
In the case of a double well with $L=10$~$\mu$m and barrier height
$E_b=6 \times 10^{-8}$~eV, we expect an axial frequency $\omega_z/2\pi=10$~MHz.

Let us give a more in-depth analysis of the dynamics of a single electron
in a double-well Penning trap. As in usual Penning traps, the radial motion
(in the $xy$ plane) and the spin motion are practically decoupled from
the axial motion. Hence, in the following we disregard the radial and
spin degrees
of freedom and consider only the electron motion along the $z$ axis.
The dynamics of this one-dimensional system is governed by the following
Schr\"{o}dinger equation
\begin{equation}\label{schreq}
i \hbar \frac{\partial \Psi(z,t)}{\partial t}=
-\frac{\hbar^2}{2 m}\frac{\partial^2 \Psi(z,t)}{\partial z^2}+eV(z)\Psi ,
\end{equation}
where $\Psi(z,t)$ is the axial wave function
of an electron in the electrostatic potential $V(z)$,
produced by the trap electrodes.

\begin{figure}
  \includegraphics[width=\columnwidth]{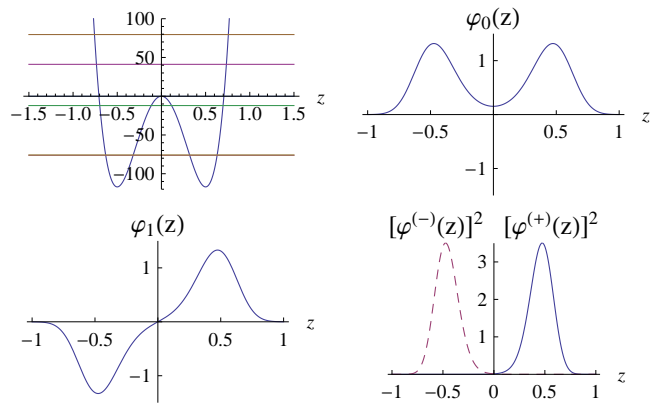}
 \caption{\label{sin2}
The panels show:
the double well potential energy, Eq.~(\ref{funct}), with the first few
energy levels,
the probability amplitude of the first eigenstate $\varphi_0$,
the probability amplitude of the second eigenstate $\varphi_1$, and the
probability density of the localized states $\varphi^{(\pm)}$.
The two lowest energy eigenvalues
appear indistinguishable in the plot.
Energies and lengths are expressed, respectively, in units of $\hbar^2/2mL^2$
and $L$.}
\end{figure}

To determine the evolution of $\Psi (z,t)$ we have to find eigenvalues and
eigenstates of the system Hamiltonian by solving the time independent
Schr\"{o}dinger equation
\begin{equation}\label{indeq}
-\frac{\hbar^2}{2m_e}\frac{d^2 \varphi(z)}{d z^2}+eV(z)\varphi(z)=E\varphi(z).
\end{equation}
Let us indicate with $\{\varphi_0,\varphi_1,\ldots\}$ the set of energy
eigenstates and with $\{E_0,E_1,\ldots\}$ the corresponding eigenvalues
solutions of Eq. (\ref{indeq}). As a consequence of the double well shape,
the energy eigenvalues appear as a series of pairs of nearby values.
We focus on the two eigenstates corresponding to the lowest energy eigenvalues,
that we suppose being smaller than the height of the barrier between the two
wells. The higher the barrier the smaller the energy difference between
these levels. In the limit of an infinitely high barrier,
these two energy eigenstates tend to become degenerate.
Because of the symmetry of the potential energy, eigenstates have a definite
parity: the eigenstate $\varphi_0$ is even and has zero nodes,
whereas the eigenstate $\varphi_1$ is odd and has one node
(see Fig.~\ref{sin2}).

Given the pair of eigenstates $\varphi_0$ and $\varphi_1$ we can construct
two orthogonal states $\varphi^{(+)}$ and $\varphi^{(-)}$ defined as
\begin{equation}
\varphi^{(\pm)}=\frac{1}{\sqrt{2}}(\varphi_0 \pm \varphi_1).
\end{equation}
An electron prepared in the state $\varphi^{(+)}$
[$\varphi^{(-)}$] is localized in the right (left) well, as displayed
in Fig.~\ref{sin2}.
However, since the states $\varphi^{(\pm)}$ are not stationary,
the particle tunnels through the barrier,
oscillating back and forth between the two wells.
For a particle initially localized in the right well,
the state of the system at a later time $t$ is described by
\begin{equation}
\Psi(z,t)=\frac{1}{\sqrt{2}} \left[
        \varphi_{0}(z)e^{-i\frac{E_{0}t}{\hbar}}
     +  \varphi_{1}(z)e^{-i\frac{E_{1}t}{\hbar}}
     \right] .
\end{equation}
Hence, the probability density to find the electron in the $z$ direction is
\begin{equation} \label{osc_1}
|\Psi(z,t)|^2=\frac{1}{2}[\varphi_{0}^2(z)+\varphi_{1}^2(z)
             +2\varphi_{0}(z)\varphi_{1}(z)\cos(\omega_{1,0} t)] ,
\end{equation}
where $\omega_{1,0}\equiv (E_{1}-E_{0})/\hbar$.
In terms of the states $\varphi^{(\pm)}$ we can recast Eq.~(\ref{osc_1}) as
\begin{equation}\label{osc}
|\Psi(z,t)|^2 = \left[ \varphi^{(+)} \right]^2
                \cos^2\left(\frac{\omega_{1,0} t}{2}\right)
             + \left[ \varphi^{(-)} \right]^2
                \sin^2\left(\frac{\omega_{1,0} t}{2}\right).
\end{equation}
From Eq.~(\ref{osc}) we clearly see that the electron oscillates between the
two wells at the tunneling frequency
$\omega_{1,0}/2 \pi$.

\begin{figure}[t]
  \includegraphics[width=\columnwidth]{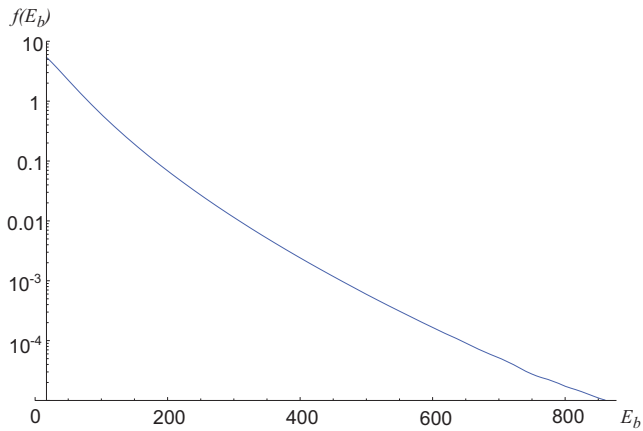}
  \caption{\label{funct2} Plot of the numerically
              calculated dimensionless quantity $f(E_b)$
              as a function of the barrier height
              $E_b$ measured in units of $\hbar^2 / 2mL^2$.}
\end{figure}

Let us consider a double-well potential energy described by the
polynomial function of Eq.~(\ref{funct}).
We have numerically calculated
the energy eigenstates and eigenvalues for different
values of the energy barrier $E_b$ and of the distance $L$ between the
two wells.
From this numerical analysis, we have deduced the following
relation
\begin{equation} \label{tunnel}
\frac{\omega_{1,0}}{2 \pi}=\frac{f(E_b)}{4 \pi m L^2}
=\frac{f(E_b)}{L^2} \times 9.18 \times 10^{-6} \, \mbox{Hz m}^2,
\end{equation}
which expresses the dependence of the tunneling frequency upon
the double well parameters.
The behavior of the dimensionless function $f$ is represented in
Fig.~\ref{funct2}.
Given the barrier height $E_b$ and width $L$,
one can calculate the corresponding
tunneling frequency by means of Eq.~(\ref{tunnel}).
\begin{figure}[t]
  \includegraphics[width=\columnwidth]{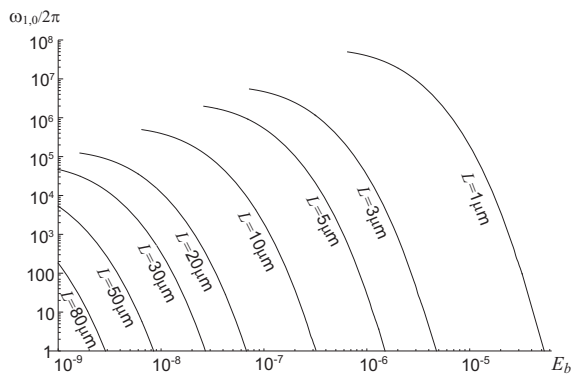}
  \caption{\label{loglog} The tunneling frequency
          $\omega_{1,0}/2\pi$ (in Hz) as a function of the barrier height
          (in eV) for different values of the well distance.
          Notice that, for any given well distance, there is
          a minimum value of the energy barrier, below which
          the wells cannot longer accomodate
          the two lowest energy eigenvalues.
          In this case we cannot define a tunneling frequency $\omega_{1,0}$.}
\end{figure}
In Fig.~\ref{loglog} it is plotted the tunneling frequency
$\omega_{1,0}/2\pi$ as a function of the double well barrier height,
for different choices of the well distance $L$.

In particular, for a double well with a distance betweeen the minima
$L=10$~$\mu$m and
a barrier height $E_b=6 \times 10^{-8}$~eV the tunneling frequency for
an electron is $\omega_{1,0}/2\pi=50$~kHz.
This value is achievable with an optimized mirror-image planar
Penning trap
of the kind proposed by Ref.~\cite{goldman},
whose central disc electrode has a radius of 100~$\mu$m.
Also the required control and stability of the applied voltages
is within the reach of present technology.
Hence, it seems feasible to employ the same physical device
to
produce both a perfectly harmonic axial potential and a double-well
one.
This possibility opens new perspectives and applications for a
single trapped electron in a planar Penning trap.
The observation of the electron oscillations between the two
wells could provide insight into the coherence properties
of the system
and measure its sensitivity to technical imperfections
and environmental noise.

\end{document}